\documentclass[aps,pra,onecolumn,superscriptaddress,showpacs,notitlepage,float,nofootinbib]{revtex4-2}
\usepackage{graphicx} 

\usepackage{centernot}
\usepackage[utf8]{inputenc}
\usepackage{amsmath}
\usepackage{empheq}
\usepackage{amssymb}
\usepackage{mathbbol}
\usepackage{latexsym}
\usepackage{ifpdf}
\usepackage{slashed}
\usepackage{color}
\usepackage{mathtools}
\usepackage{soul}
\usepackage{hyperref}
\usepackage{mathrsfs}

\begin{document}

\title{Scattering Faddeev calculations in the double continuum}
\author{Romain Guérout}
\affiliation{Université Paris-Saclay, CNRS, Laboratoire Aimé Cotton, 91405, Orsay, France.}

\begin{abstract}
    We use the configuration-space Faddeev formalism to study scattering of three particles in the double continuum where all particles are free. All scattering processes, starting from and ending in both single and double continua, are collected in a unique matrix. We apply our method to the benchmark system of neutron-deuteron scattering.
\end{abstract}

\maketitle

\section{Introduction}

The Faddeev formalism~\cite{fadeev1961scattering} describes the quantum mechanical scattering of three particles. It has been developed to circumvent the failure of the Lippmann-Schwinger equation to give a unique solution to the Schrödinger equation. When working in configuration-space, a unique solution is obtained after imposing proper boundary conditions on the wavefunction. For bound state calculations as well as the scattering of one particle on the bound cluster of the other two, those boundary conditions are well known. For situations where all three particles are free, a few years were needed to carefully derive those boundary conditions~\cite{merkuriev1976three}.

Configuration-space Faddeev scattering calculations in the double continuum are now routinelly performed. Notably, we can quote calculations for the three helium atoms system at low energy~\cite{kolganova1998three} and importantly the scattering from a neutron or a proton on a deuteron which has since become a benchmark~\cite{friar1995benchmark} system for a variety of different theoretical methods.

The difficulty in treating the double continuum correctly lie in the fact that the caclulated wavefunction contains both the information about two-particle channels and three-body breakup channels. Disantangling those two behaviours from the wavefunction presents some difficulties since those channels are not orthogonal in the usual sense (we note however the elegant method presented in~\cite{belov2013asymptotic} where those two contributions are indeed orthogonalized).

Here, we report on a simple method to extract both the two-body and the three-body contributions from the wavefunction which stems from the fact that those two different contributions are well describe in two different coordinate systems : Jacobi coordinates for the two-body channels and polar coordinates for the three-body breakup channels. We then apply our methods to the neutron-deuteron scattering.

\section{Faddeev formalism}

In this work, we are interested in the scattering processes between three particles. Incoming and/or outgoing states of the three particles can be of two types. It can be a state where one of the particle is free with respect to a bound cluster of the other two particles. This state will be refered to as a $(1+2)$ or single continuum state. The other type of asymptotic state is one where all three particles are free with respect to each other. We will refer to that state as a $(1+1+1)$ or double continuum state.

The origin of the terms single or double continuum states is as follow. In a $(1+2)$ state, the energy between the free particle and the cluster is unrestricted and can take any value. It forms a continuum. In a $(1+1+1)$ state, the total energy is also unrestricted and forms a continuum. In that state though, there is another continuous degree of freedom which is how this total energy is partitioned between the internal coordinates of the particles. This partition forms another continuum.

For three particles, there are two internal coordinates which can be taken as the Jacobi vectors $\mathbf{x}$ and $\mathbf{y}$ . While the first continuum will be of course the total energy $E$, the partition of this total energy among the Jacobi vectors can be parametrized by the ratio $y/x$ of the Jacobi vectors norms which will then constitutes our second continuum.

We use mass-scaled Jacobi coordinates $\{\textbf{x}_i,\textbf{y}_i\}$ and $(i,j,k)$ is a cyclic permutation of $(1,2,3)$ (we take $\hbar = 1$ in the following)
\begin{align}
    \textbf{x}_i &= \tau_{\textbf{x}_i} \left( \textbf{r}_j - \textbf{r}_k \right) \\
    \textbf{y}_i &= \tau_{\textbf{y}_i} \left( \textbf{r}_i - \frac{m_j \textbf{r}_j + m_k \textbf{r}_k}{m_j + m_k} \right)
\end{align}
where $\tau_{\textbf{x}_i} = \sqrt{2 \mu_{jk}}$, $\tau_{\textbf{y}_i} = \sqrt{2 \mu_{i,jk}}$, $\mu_{jk} = \frac{m_j m_k}{m_j + m_k}$ and $\mu_{i,jk} = \frac{m_i(m_j + m_k)}{m_i + m_j + m_k}$. The $\tau$ factors absorb the masses so that the kinetic energy operator is $\hat{T}_i = -\Delta_{\textbf{x}_i} - \Delta_{\textbf{y}_i}$. We remark that
\begin{equation}
    \tau_{\textbf{x}_i}\tau_{\textbf{y}_i} = 2 \sqrt{\frac{\prod_{n} m_n}{\sum_{n} m_n}} = 2 \mu_{3B}
\end{equation}
where $\mu_{3B}$ is the three-body reduced mass which is independent of the Jacobi arrangement (more generally, the $N$-body reduced mass is $\mu_{NB}^{N-1} = \frac{\Pi m}{\Sigma m}$). The Jacobi vector $\textbf{x}_i$ is represented in spherical coordinates consisting of its norm $x_i$ and spherical angles $\hat{\textbf{x}}_i \equiv (\theta_{\textbf{x}_i},\varphi_{\textbf{x}_i})$. The same goes for $\textbf{y}_i$. The 6D vector $(\textbf{x}_i,\textbf{y}_i)$ can therefore be written as $(x_i,y_i,\hat{\textbf{x}}_i,\hat{\textbf{y}}_i)$ corresponding to the partition $\mathbb{R}^6 = \mathbb{R}^2 \otimes \mathbb{R}^4$ into two radii and four angles. We introduce the polar coordinates~\footnote{Those polar coordinates are closely related to the Delves coordinates which use $\alpha_i=\text{arctan}(x_i/y_i)$} $(\rho,\alpha_i,\hat{\textbf{x}}_i,\hat{\textbf{y}}_i)$ as
\begin{align}
    \rho^2 &= x_i^2 + y_i^2 \\
    \alpha_i &= \text{arctan}(y_i/x_i)
\end{align}
In the following, we might refer to Jacobi coordinates as cartesian for obvious reasons. Note that the hyperradius $\rho$ is independent of the Jacobi arrangement.

The total wavefunction $\Psi$ is the solution of the Schrödinger equation $\left( \hat{T} + \hat{V} - E \right) \Psi = 0$. The zero for the center of mass energy $E$ is taken to be at breakup. Therefore, above-breakup scattering means that $E \ge 0$. The potential term is taken as a sum of pairwise, central potentials \textit{i.e.} $V = V_1(x_1) +V_2(x_2) +V_3(x_3)$. The total wavefunction $\Psi$ is decomposed into three Faddeev components $\phi_i$ satisfying three coupled equations
\begin{align}
    \Psi(\textbf{x}_i,\textbf{y}_i) &= \phi_1(\textbf{x}_1,\textbf{y}_1) + \phi_2(\textbf{x}_2,\textbf{y}_2) + \phi_3(\textbf{x}_3,\textbf{y}_3) \label{faddeevDecomp} \\
    \left( \hat{T}_i + \hat{V}_i - E \right) \phi_i(\textbf{x}_i,\textbf{y}_i) &= -\hat{V}_i \left[ \phi_j(\textbf{x}_j,\textbf{y}_j) + \phi_k(\textbf{x}_k,\textbf{y}_k) \right] \label{totalHamilt}
\end{align}
The 4D angular part is solved through partial waves expansion with respect to bipolar spherical harmonics $\mathcal{Y}_{a}(\hat{\textbf{x}}_i,\hat{\textbf{y}}_i)$ defining radial wavefunctions $f_{i,a}(x_i,y_i)$ for component $i$ in partial wave $a$ as
\begin{subequations}
    \begin{align}
        \phi_i(\textbf{x}_i,\textbf{y}_i) &= \sum_{a} \frac{f_{i,a}(x_i,y_i)}{x_i y_i} \mathcal{Y}_{a}(\hat{\textbf{x}}_i,\hat{\textbf{y}}_i) \\
        \mathcal{Y}_{a}(\hat{\textbf{x}}_i,\hat{\textbf{y}}_i) &= \left[ Y_{\ell_a}(\hat{\textbf{x}}_i) \otimes Y_{\lambda_a}(\hat{\textbf{y}}_i) \right]_{LM} \\
        \hat{\ell}_{\textbf{x}_i}^{2} \mathcal{Y}_{a}(\hat{\textbf{x}}_i,\hat{\textbf{y}}_i) &= \ell_a(\ell_a + 1) \mathcal{Y}_{a}(\hat{\textbf{x}}_i,\hat{\textbf{y}}_i) \\
        \hat{\ell}_{\textbf{y}_i}^{2} \mathcal{Y}_{a}(\hat{\textbf{x}}_i,\hat{\textbf{y}}_i) &= \lambda_a(\lambda_a + 1) \mathcal{Y}_{a}(\hat{\textbf{x}}_i,\hat{\textbf{y}}_i)
    \end{align}
\end{subequations}
For simplicity, we consider partial waves consisting of only orbital angular momentum channels \emph{i.e.} bipolar spherical harmonics. In general, they can also include spin angular momentum. This will be illustrated later when we will apply our formalism to neutron-deuteron scattering.

The total wavefunction $\Psi$ can itself be expanded in partial waves as 
\begin{equation}
    \Psi(\textbf{x}_i,\textbf{y}_i) = \sum_{a} \frac{u_{a}(x_i,y_i)}{x_i y_i} \mathcal{Y}_{a}(\hat{\textbf{x}}_i,\hat{\textbf{y}}_i)
\end{equation}
defining radial wavefunction $u_a(x_i,y_i)$. Inserting those partial waves expansions in equation~\eqref{faddeevDecomp} and use the orthogonality relation for the bipolar spherical harmonics allows to get an expression for the total wavefunction radial part $u_a$ in terms of the component radial parts $f_{i,a}$
\begin{align}
    u_{a}(x_i,y_i) &= f_{i,a}(x_i,y_i) + \sum_{b} \iint d\hat{\textbf{x}_i} d\hat{\textbf{y}_i} \frac{x_i y_i}{x_j y_j} \mathcal{Y}_{a}^{*}(\hat{\textbf{x}}_i,\hat{\textbf{y}}_i) f_{j,b}(x_j,y_j) \mathcal{Y}_{b}(\hat{\textbf{x}}_j,\hat{\textbf{y}}_j) \nonumber \\
    & + \sum_{c} \iint d\hat{\textbf{x}_i} d\hat{\textbf{y}_i} \frac{x_i y_i}{x_k y_k} \mathcal{Y}_{a}^{*}(\hat{\textbf{x}}_i,\hat{\textbf{y}}_i) f_{k,c}(x_k,y_k) \mathcal{Y}_{c}(\hat{\textbf{x}}_k,\hat{\textbf{y}}_k) \\
    u_{a}(x_i,y_i) & \equiv f_{i,a}(x_i,y_i) + \sum_b \mathcal{J}_{ij}^{ab}\left[f_{j,b}\right](x_i,y_i) + \sum_c \mathcal{J}_{ik}^{ac}\left[f_{k,c}\right](x_i,y_i) \label{radiaPsiVsradialPhi}
\end{align}
where we have introduced the integral transform $\mathcal{J}_{ij}^{ab}$ of a generic function $h$
\begin{equation}
    \mathcal{J}_{ij}^{ab}\left[ h \right] = \iint d\hat{\textbf{x}_i} d\hat{\textbf{y}_i} \frac{x_i y_i}{x_j y_j} \mathcal{Y}_{a}^{*}(\hat{\textbf{x}}_i,\hat{\textbf{y}}_i) h(\mathbf{x}_j,\mathbf{y}_j) \mathcal{Y}_{b}(\hat{\textbf{x}}_j,\hat{\textbf{y}}_j)
\end{equation}
In equation~\eqref{radiaPsiVsradialPhi}, this integral transform takes in a radial wavefunction for component $j$ in a partial wave $b$ and outputs a radial wavefunction for component $i$ in partial wave $a$. We refer to it as Jacobi transform~\footnote{In the literature, an unrelated Jacobi transform already exists.}. It plays a central role in the Faddeev formalism since it is this term which couples the equations for the different components. It is also the tool which allows to sum up the contribution from the different Faddeev components to get physical observables. We note that in polar coordinates, it is independent of the hyperradius $\rho$ : given a function of $\alpha_j$, it then outputs a function of $\alpha_i$.

Inserting the partial wave expansion in equation~\eqref{totalHamilt} leads to 2D coupled integro-differential equations for the radial wavefunctions $f_{i,a}$
\begin{equation}
    \left( t_{i,a} - E \right) f_{i,a}(x_i,y_i) + \sum_{b} v_{i,ab} \, f_{i,b}(x_i,y_i) = -\sum_{b,c} v_{i,ac} \sum_{j \neq i} \mathcal{J}_{ij}^{cb}\left[ f_{j,b} \right](x_i,y_i) \label{componentEq}
\end{equation}
where $t_{i,a} = -\frac{\partial^2}{\partial x_i^2} - \frac{\partial^2}{\partial y_i^2} + \frac{\ell_a(\ell_a + 1)}{x_i^2} + \frac{\lambda_a(\lambda_a + 1)}{y_i^2}$ and 
\begin{equation}
    v_{i,ab} = \mathcal{J}_{ii}^{ab} \left[ V_i \right] = \iint d\hat{\textbf{x}_i} d\hat{\textbf{y}_i} \mathcal{Y}_{a}^{*}(\hat{\textbf{x}}_i,\hat{\textbf{y}}_i) V_i(\textbf{x}_i,\textbf{y}_i) \mathcal{Y}_{b}(\hat{\textbf{x}}_i,\hat{\textbf{y}}_i)
\end{equation}
For central isotropic pairwise potential $V_i(x_i)$, we obviously have $v_{i,ab} = V_i(x_i) \delta_{ab}$.

To solve numerically the equation~\eqref{componentEq}, the partial Faddeev components $f_{i,a}$ are expanded onto a basis of cubic Hermite splines $s$
\begin{equation}
    f_{i,a}(x_i,y_i) = \sum_{n,m} a_{nm}^{i,a} \, s_{n}(x_i) \, s_{m}(y_i)
\end{equation}
The numerical solution of equation~\eqref{componentEq} consists in finding the expansion coefficients $a_{nm}^{i,a}$. Specifying proper grids in $x_i$ and $y_i$ and use an orthogonal collocation method leads to the creation of the different operators appearing in equation~\eqref{componentEq} : the kinetic energy operator $\mathbb{T}$, an indicator operator $\mathbb{1}$, a potential operator $\mathbb{V}$ and a Jacobi kernel operator $\mathbb{K}$. While the kinetic and potential operators are self-explanatory, the indicator operator $\mathbb{1}$ simply transforms from expansion coefficients to values at the collocation points (it accounts for the energy $E$ in equation~\eqref{componentEq}) and the kernel operator $\mathbb{K}$ couples the different equations and therefore implements the Jacobi transform. Explicit expressions for the matrix elements of those operators can be found in~\cite{guerout2024calculation}.

A solution of equation~\eqref{componentEq} involves the resolution of the inhomogeneous linear system $\left( \mathbb{T} + \mathbb{V} + \mathbb{K} - E \mathbb{1} \right)v = - \mathbb{K} \chi$ where the vector $\chi$ contains the information about the incoming state as well as enforces the boundary conditions. The solution vector $v$ contains the expansion coefficients $a_{nm}^{i,a}$. We will go into more details about the numerical methods we use in the following sections.

\section{Asymptotic channels functions}

\subsection{(1+2) asymptotic channels}

Below breakup, at $E < 0$, the only scattering processes possible are $(1+2) \to (1+2)$. The asymptotic region consists of $y_i \to \infty$ while the coordinate $x_i$ stays bounded by the potential $V_i$. The uncoupled differential equation for the Faddeev component $f_i$ is 
\begin{equation}
    \left( -\frac{\partial^2}{\partial x_i^2} -\frac{\partial^2}{\partial y_i^2} + \frac{\ell(\ell + 1)}{x_i^2} + \frac{\lambda(\lambda + 1)}{y_i^2} + V_i(x_i) - E \right) f_i(x_i,y_i) = 0
\end{equation}
This equation is separable and a general solution is
\begin{equation}
    \label{atomDimerAsympt}
    f_i(x_i,y_i) = \sum_v c_v \varphi_{v}(x_i) h^{(+)}(q y_i)
\end{equation}
where $c_v$ are complex coefficients, $\varphi_{v}(x_i)$ is a bound rovibrational wavefunction for the potential $V_i(x_i)$ with vibrational number $v$, angular momentum $\ell$ and energy $\epsilon_{v,\ell} < 0$ and $h^{(+)}(q y_i)$ is an outgoing Ricatti-Hankel function with momentum $q$ and angular momentum $\lambda$. Conservation of energy gives
\begin{equation}
    E = \epsilon_{v,\ell} + \frac{q^2}{2 \mu_{i,jk}}
\end{equation}
The sum in equation~\eqref{atomDimerAsympt} runs over vibrational numbers for which $q^2 > 0$ \textit{i.e.} over the open channels : indeed, only for real $q$ does the function $h^{(+)}(q y_i)$ carry probability flux at infinity and will then contribute to the scattering process.

The asymptotic channels functions for the $(1+2)$ scattering processes are then the functions $\varphi_{v}(x_i) h^{(+)}(q y_i)$ which we will refer to as "bound plane waves". The degrees of liberty in equation~\eqref{atomDimerAsympt} are the coefficients $c_v$ which will become the relevant scattering matrix elements once the calculated Faddeev component is matched to its asymptotic form.

\subsection{(1+1+1) asymptotic channels}
\label{subs:doubleCont}

Above breakup, at $E > 0$, the scattering process $(1+2) \to (1+1+1)$ \textit{i.e.} breakup, becomes energetically possible. The asymptotic region now consists of both $x_i \to \infty$ and $y_i \to \infty$. It now explores a region where $V_i(x_i) \to 0$ so that the potential can be neglected in the asymptotically uncoupled differential equation for $f_i$. This uncoupled differential equation is best solved in polar coordinates and reads
\begin{equation}
    \label{fullEq}
    \left( -\frac{\partial^2}{\partial \rho^2} -\frac{1}{\rho} \frac{\partial}{\partial \rho} -\frac{1}{\rho^2} \frac{\partial^2}{\partial \alpha_i^2} + \frac{\ell(\ell + 1)}{\rho^2 \cos^2 \alpha_i} + \frac{\lambda(\lambda + 1)}{\rho^2 \sin^2 \alpha_i} - E \right) f_i(\rho,\alpha_i) = 0
\end{equation}
Separation of variables posing $f_i(\rho,\alpha_i) = p(\rho) q(\alpha_i)$ leads to
\begin{subequations}
\begin{align}
    \left( -\rho^2 \frac{\partial^2}{\partial \rho^2} -\rho \frac{\partial}{\partial \rho}    - \rho^2 E \right) p(\rho) &= -\nu^2 p(\rho) \label{Prho} \\
    \left( -\frac{\partial^2}{\partial \alpha_i^2} + \frac{\ell(\ell + 1)}{\cos^2 \alpha_i} + \frac{\lambda(\lambda + 1)}{\sin^2 \alpha_i} \right) q(\alpha_i) &= \nu^2 q(\alpha_i) \label{Aalpha}
\end{align}
\end{subequations}
The equation~\eqref{Prho} is the standard Bessel differential equation and an equation closely related to~\eqref{Aalpha} appears in an early paper from Delves~\cite{delves1960tertiary}. The separation constant $\nu^2$, eigenvalue of the differential operator $-\frac{\partial^2}{\partial \alpha_i^2} + \frac{\ell(\ell + 1)}{\cos^2 \alpha_i} + \frac{\lambda(\lambda + 1)}{\sin^2 \alpha_i}$, takes values $\nu = \ell + \lambda + 2(n+1)$ with $n = 0,1,2 ...$ while the functions $q(\alpha_i)$ are the so-called Delves functions $D_{\nu}^{(\ell,\lambda)}$
\begin{equation}
    q(\alpha_i) \equiv D_{\nu}^{(\ell,\lambda)}(\alpha_i) \propto \cos^{\ell + 1} (\alpha_i) \sin^{\lambda + 1} (\alpha_i) P_{n}^{(\lambda + 1/2, \ell + 1/2)}(\cos 2 \alpha_i)
\end{equation}
where $P_n^{(a,b)}$ is a Jacobi polynomial.
We note that the Delves functions are eigenfunctions of the Jacobi transform with eigenvalues $\beta$ \textit{i.e.}
\begin{equation}
    \mathcal{J}_{ij}^{aa}\left[ D_{\nu}^{(\ell_{a},\lambda_{a})} \right] = \beta_{ij}^{a} \, D_{\nu}^{(\ell_{a},\lambda_{a})}
\end{equation}
More generally, the Jacobi transform $\mathcal{J}_{ij}^{ab}$ of a Delves function $D_{\nu}^{(\ell_b,\lambda_b)}$ is proportional to a Delves function $D_{\nu'}^{(\ell_a,\lambda_a)}$.

A general solution of equation~\eqref{fullEq} is then
\begin{equation}
    \label{asymptDelves1}
    f_i(\rho,\alpha_i) = \sum_{\nu} c_{\nu} D_{\nu}^{(\ell,\lambda)}(\alpha_i) H^{(+)}_{\nu}(k \rho)
\end{equation}
with $k = \sqrt{E}$ and $H^{(+)}_{\nu}$ an outgoing Hankel function of order $\nu$.
At this point, we can perform an additional manipulation on the functions in equation~\eqref{asymptDelves1} : unlike the previous channels functions in equation~\eqref{atomDimerAsympt}, the above sum involves outgoing Hankel functions of a unique argument $k \rho$. We can therefore use the asymptotic behavior of the Bessel functions to write everything in terms of, for example, $H_0^{(+)}$
\begin{align}
    f_i(\rho,\alpha_i) \xrightarrow[\rho \to \infty]{} &\left( \sum_{\nu} c_{\nu} e^{-i \frac{\nu \pi}{2}} D_{\nu}^{(\ell,\lambda)}(\alpha_i) \right) H^{(+)}_{0}(k \rho) \nonumber \\
    & \equiv A(\alpha_i) H^{(+)}_{0}(k \rho) \label{asymptDelves2}
\end{align}
In equation~\eqref{asymptDelves1}, the degrees of liberty are the coefficients $c_{\nu}$ while in equation~\eqref{asymptDelves2} the only degree of liberty is the unknown function $A(\alpha_i)$. Equations~\eqref{asymptDelves1} and~\eqref{asymptDelves2} are two different interpretations of the fact that a $(1+1+1)$ asymptotic channel lie in a double continuum. It must then be described either by an (countably) infinite number of scalar scattering matrix elements or by a single function of the continuous variable $\alpha_i$.

Here, we adopt the point of view of equation~\eqref{asymptDelves2} : the asymptotic channels functions for the $(1+1+1)$ scattering processes are the functions $H^{(+)}_{0}(k \rho)$ which we will refer to as "cylindrical waves". The matching of the calculated Faddeev component to its asymptotic form will lead to scattering matrix elements which will be the breakup amplitudes, functions of the variable $\alpha_i$.

\subsection{Reaching the asymptotic region}
\label{sec:reaching}

Reaching the asymptotic region does not occur on the same footing for $(1+2)$ single continuum channels and the $(1+1+1)$ double continuum channels. For this discussion, we assume that the various pairwise potentials $V_i(x_i)$ have a finite range $b$ (this range can be, for instance, the region for which the least bounded two-body wavefunction have significant values) and are short-range (thus we exclude the Coulomb potential which require a particular treatment). We note that the coupling term in the right-hand side of equation~\eqref{componentEq} involves the potential $V_i(x_i)$ and the other partial Faddeev components $f_j(x_j,y_j)$ and $f_k(x_k,y_k)$.

The single continuum channel $(1+2)$ asymptotic region for the partial component $f_i(x_i,y_i)$ corresponds to $x_i < b$ and $y_i \to \infty$. In this limit, it is easy to see that both $x_j$ and $x_k$ tends towards $y_i$. Therefore, both partial components $f_j$ and $f_k$ will vanish as they will involve two-body wavefunctions in the potential $V_j$ and $V_k$ evaluated at $x_j, x_k \to \infty$. Therefore, the $(1+2)$ asymptotic region is reached for $y_i \gg b$. We note that all the Jacobi transforms therefore vanish in equation~\eqref{radiaPsiVsradialPhi} and that the Faddeev component actually becomes the total wavefunction $\phi_i(\textbf{x}_i,\textbf{y}_i) \to \Psi(\textbf{x}_i,\textbf{y}_i)$.

The situation is quite different for the double continuum channel $(1+1+1)$. Here, all components $f$ have non-negligible values everywhere in the $(x,y)$ plane. It is not obvious at first whether the coupling term even vanishes. A careful analysis~\cite{glockle1992boundary} shows that the coupling term in this case actually behaves as $1/y^{3/2}$ for $y \to \infty$ and $x$ finite. So, the asymptotic region can be eventually reached but much slower than in the single continuum case as it results from interferences between the three Faddeev partial components. Additionally, for a given $y_{max}$ we can define a critical polar angle $\alpha_c = \text{arctan} \left(b \tau_y/y_{max}\right)$. The effect of the potential will have a lingering effect on the breakup amplitudes for $\alpha_c < \alpha < \pi/2$. This is the region where two of the particles stay close to each other for a very long time. It originates from the fact that very little of the total energy $E$ has been partitioned in the $x$ degree of liberty. Only for $y_{max} \to \infty$ does $\alpha_c \to \pi/2$. For this reason, calculations done in a finite box $\{0<x<x_{max},0<y<y_{max}\}$ will usually have breakup amplitudes  imprecise near $\alpha \approx \pi/2$ and appropriate methods~\cite{belov2013asymptotic} are needed to actually retrieve the breakup amplitude at infinity.

\subsection{Structure of the scattering matrix above breakup}

At $E > 0$, the $(1+1+1)$ channel is open so that those scattering processes are possible
\begin{subequations}
\begin{align}
    (1+2) & \to (1+2) & \text{elastic }(1+2) \\
    (1+2) & \to (1+1+1) & \text{breakup} \\
    (1+1+1) & \to (1+2) & \text{3B recombination} \\
    (1+1+1) & \to (1+1+1) & \text{elastic }(1+1+1)
\end{align}
\end{subequations}

In practice, we rather calculate the transition matrix $\mathbf{T}$. Schematically, and of we collect all the $n$ $(1+2)$ channels in a first block and all the $p$ $(1+1+1)$ channels in a second block, solving the scattering problem amounts at finding a set of $N=n+p$ (the number of open channels) linearly independent solutions $(f^1,f^2,...,f^N)$ of the coupled Faddeev equations which behaves asymptotically as
\begin{equation}
    \label{scattEq}
    \left( \begin{array}{c}
        f^1 \\
        \vdots \\
        f^n \\
        \hline
        f^{n+1} \\
        \vdots \\
        f^N
    \end{array} \right)
    \approx 
    \left( \begin{array}{c|c}
        \begin{matrix} 1 & & & \\  & \ddots & \\ & & 1 \end{matrix} & 0 \\
         \hline
        0 & \begin{matrix} i_1(\alpha) & & & \\  & \ddots & \\ & & i_p(\alpha) \end{matrix}
    \end{array} \right)
    \left( \begin{array}{c}
        \varphi_{1}(x) j(q_1 y) \\
        \vdots \\
        \varphi_{n}(x) j(q_n y) \\
        \hline
        J_{0}(k \rho)\\
        \vdots \\
        J_{0}(k \rho)
    \end{array} \right) +
    \left( \begin{array}{c|c}
        \begin{matrix} & & & \\  & T_{11} & \\ & & \end{matrix} & \begin{matrix} & & & \\  & T_{12}(\alpha) & \\ & & \end{matrix} \\
         \hline
        \begin{matrix} & & & \\  & T_{21} & \\ & & \end{matrix} & \begin{matrix} & & & \\  & T_{22}(\alpha) & \\ & & \end{matrix} 
    \end{array} \right)
    \left( \begin{array}{c}
        \varphi_{1}(x) h^{(+)}(q_1 y) \\
        \vdots \\
        \varphi_{n}(x) h^{(+)}(q_n y) \\
        \hline
        H^{(+)}_{0}(k \rho)\\
        \vdots \\
        H^{(+)}_{0}(k \rho)
    \end{array} \right)
\end{equation}
The functions $i_1(\alpha)$,...,$i_p(\alpha)$ characterize the incoming $(1+1+1)$ channels and the functions $j$ and $J_{0}$ are regular (Ricatti)-Bessel functions. The blocks $T_{11}$ (elastic) and $T_{21}$ (3B recombination) contains scalar complex numbers while the blocks $T_{12}$ (breakup) and $T_{22}$ (elastic) consist of complex functions of the polar angle $\alpha$. Transition and scattering matrices are related via $\mathbf{S} = \mathbf{1} + 2 i \mathbf{T}$. We will illustrate in the next section how to retrieve the usual unitarity and reciprocity properties from the scattering matrix $\mathbf{S}$.

\section{Application : the neutron-deuteron scattering}

We apply the Faddeev formalism to the neutron ($n$) scattering on the deuteron ($d$). This system has become a benchmark test for various numerical methods~\cite{friar1995benchmark}. The different scattering processes we will describe are
\begin{subequations}
\begin{align}
    n + d & \to n + d & \text{elastic } nd \\
    n + d & \to n + n + p & \text{breakup} \\
    n + n + p & \to n + d & \text{3B recombination} \\
    n + n + p & \to n + n + p & \text{elastic } nnp
\end{align}
\end{subequations}

We adopt the same level of approximation as in the benchmarks and consider a system of three identical nucleons. Proton (p) and neutron only differ by their isospin state. Therefore, there is a unique Faddeev component $\phi$ to be calculated. The nucleon mass $m$ is set to $\hbar^2 / m = 41.47$ MeV fm$^2$. We calculate for the doublet system $J^{\Pi}=1/2^{+}$ and consider the following two channels written in a $LS$ coupling scheme :
\begin{table}[h!]
    \centering
    \begin{tabular}{ccccc}
    \hline
    \hline
    channel $a$ & ($\ell_{a}$,$\lambda_{a}$)$L$ & ($s_{a}$,$\sigma_{a}$)$S$  & $t_{a}$ & name \\
    \hline \\[-2ex]
    1 & (0,0)0 & (0,$\frac12$)$\frac12$ & 1 & singlet \\[1ex]
    2 & (0,0)0 & (1,$\frac12$)$\frac12$ & 0 & triplet \\[0.5ex]
    \hline
    \hline
    \end{tabular}
    \caption{The two channels for the doublet $(J^{\Pi} = 1/2^{+})$ system.}
    \label{tab:channels}
\end{table}

In those channels, $s_{a}$ ($t_{a}$) refer to the (iso)spin of a pair while $\sigma_{a}$ is the spin of the spectator. In the evaluation of the kernel operator which acts between different Jacobi arrangements, recoupling coefficients of the form $\langle (s_j s_k)s_{jk},s_i,S | (s_i s_k)s_{ik},s_j,S \rangle$ have to be considered for both the spins and the isospins. Standard expressions for those recoupling coeffcients in terms of 6-\textit{j} symbols~\cite{edmonds1996angular} lead to the following spin $w_s$ and isospin $w_t$ recoupling matrices in the basis of the two channels above
\begin{subequations}
    \begin{align}
    w_s &= \begin{pmatrix}
        -\frac12 & -\frac{\sqrt{3}}{2} \\[1ex]
        \frac{\sqrt{3}}{2} & -\frac12
    \end{pmatrix} \\
    w_t &= \begin{pmatrix}
        -\frac12 & \frac{\sqrt{3}}{2} \\[1ex]
        -\frac{\sqrt{3}}{2} & -\frac12
    \end{pmatrix}
\end{align}
\end{subequations}
The total recoupling matrix $w$ is then $w = w_s \odot w_t$ :
\begin{equation}
    w = \begin{pmatrix}
        \frac14 & -\frac34 \\[1ex]
        -\frac34 & \frac14
    \end{pmatrix}
\end{equation}
where $\odot$ is the Hadamard (element-wise) product. In practice, once a kernel operator $\mathbb{k}$ has been evaluated with respect to the orbital angular momentum channel $\left( \ell_{a},\lambda_{a} \right)L = \left(0,0\right)0$ the full operator in the basis of the channels in table~\eqref{tab:channels} is $\mathbb{K} = w \otimes \mathbb{k}$ where $\otimes$ is the Kronecker product.

The pairwise potentials are taken to be sums of Yukawa potentials~\cite{friar1995benchmark}
\begin{align}
    V_1(x) &= 1438.72 \, \frac{e^{-3.11 x}}{x} - 513.968 \, \frac{e^{-1.55 x}}{x} \\
    V_2(x) &= 1438.72 \, \frac{e^{-3.11 x}}{x} - 626.885 \, \frac{e^{-1.55 x}}{x}
\end{align}
with distances expressed in fm and energies in MeV. With these parameters, the deuteron is the bound state of the triplet channel potential $V_2$ and has an energy of $E_d = -2.2306$ MeV.

To make the connection with equation~\eqref{scattEq}, our scattering matrix $\mathbf{S}$ will be at most of dimension $3 \times 3$ consisting of the unique triplet $(1+2)$ asymptotic channel and the singlet and triplet $(1+1+1)$ channels : 
\begin{equation}
    \mathbf{S} = \left( \begin{array}{c|cc}
        S_{11} & S_{12}(\alpha)  & S_{13}(\alpha) \\
         \hline
        S_{21} & S_{22}(\alpha) & S_{23}(\alpha) \\
        S_{31} & S_{32}(\alpha) & S_{33}(\alpha)
    \end{array} \right)
    \label{scatteringMatrix}
\end{equation}
Obviously, for $E < 0$ only the single continuum channel is open and the scattering matrix reduces to $\mathbf{S} = \left( S_{11} \right) $.

Finally the off-diagonal blocks of the scattering matrix above relate processes where the reduced mass changes between incoming and outgoing states. Those matrix elements have to be renormalized in order to be able to properly compare incoming and outgoing probability fluxes. It is quite clear that the reduced mass which enters the probability flux for a single continuum $(1+2)$ channel is $\mu_{i,jk}$, the reduced mass in the $y$ direction. The reduced mass to be used in a double continuum $(1+1+1)$ channel is given by the Jacobian of the transformation between cartesian and polar coordinates~\cite{nadal2011method}. This Jacobian involves the quantity $\sqrt{\mu_{jk} \mu_{i,jk}}$ which is precisely the three-body reduced mass $\mu_{3B}$. Therefore, we define a reduced mass matrix $\boldsymbol{\mu}$ as
\begin{equation}
    \boldsymbol{\mu} = \left( \begin{array}{c|cc}
        1 & \sqrt{\mu_{i,jk}/\mu_{3B}}  & \sqrt{\mu_{i,jk}/\mu_{3B}} \\
         \hline
        \sqrt{\mu_{3B}/\mu_{i,jk}} & 1 & 1 \\
        \sqrt{\mu_{3B}/\mu_{i,jk}} & 1 & 1
    \end{array} \right)
\end{equation}
and we renormalized the scattering matrix as $\mathbf{S} \to \boldsymbol{\mu} \odot \mathbf{S}$.

\subsection{Numerical methods}
\label{subs:numerical}

Calculations can be done either in cartesian or polar coordinates. Both methods have their advantages and drawbacks. Cartesian grids are well adapted to describe the bound plane wave part of the Faddeev components but the kernel operator acts on both $x$ and $y$ grids. Polar grids are obviously adapted to describe the cylindrical wave part of the components and the fact that the kernel operator only act on the $\alpha$ grid simplify its construction. While our previous work~\cite{guerout2024calculation} was done in cartesian coordinates, here we work in polar coordinates. The grid in $\rho$ is chosen to cover distances of the order of $\tau_x^{-1} x \approx \tau_y^{-1}y \approx 100$ fm.
It runs up to $\frac{\rho}{\sqrt{\tau_{x}\tau_{y}}} = 140$ fm with $n_{\rho} = 500$ collocation points.
When doing calculations in polar coordinates, care must be taken to the grid in $\alpha$ : a high density of points is needed around $\alpha \approx \pi/2$ so that the grid correctly sample the deuteron bound state wavefunction. We use the method described in~\cite{roudnev2011automatic} to construct an optimized grid in $\alpha$ with $n_{\alpha} = 256$ collocation points. Both grids in $\rho$ and $\alpha$ are non uniform with more points at small $\rho$ and for $\alpha$ near 0 and $\pi/2$. We have verified that increasing the number of points in both grids did not change significantly our results anymore.

The boundary condition imposed on the $\alpha$ grid is that the partial radial Faddeev component must vanish at $\alpha=0$ and $\alpha=\pi/2$. Operators are first calculated without imposing a boundary condition in $\rho$. The boundary condition in $\rho$ is imposed in each open channel by replacing the last two cubic spline basis functions by an appropriate linear combination so as to impose a particular logarithmic derivative. 

Below breakup, at $E<0$, the imposed boundary condition consists of that of an outgoing bound plane wave (see equation~\eqref{atomDimerAsympt}). This leads to an inhomogeneous boundary condition where the logarithmic derivative in $\rho$ depends on $\alpha$. Above breakup, we impose a boundary condition of an outgoing Hankel function (see equation~\eqref{asymptDelves2}). This boundary condition does not depend on $\alpha$. In this case, we note however that this boundary condition is not adapted to the bound plane wave part of the partial Faddeev component : while the calculated component will consist mostly of an outgoing Hankel function $H^{(+)}_{0}(k \rho)$, the bound plane wave part of it will be a superposition of regular and outgoing functions.

A solution is obtained by solving the following linear system
\begin{subequations}
\label{operatorsEq}
\begin{align}
    \left( \mathbb{T} + \mathbb{V} + \mathbb{K} - E \mathbb{1} \right)v &= - \mathbb{K} \chi & \text{for a } (1+2) \text{ incoming state} \\
    \left( \mathbb{T} + \mathbb{V} + \mathbb{K} - E \mathbb{1} \right)v &= - (\mathbb{V} + \mathbb{K}) \chi & \text{for a } (1+1+1) \text{ incoming state} 
\end{align}
\end{subequations}
where the vector $\chi$ contains expansion coefficients of the appropriate incoming state. As shown in equation~\eqref{scattEq} this incoming state is either a bound plane wave (a solution of $\left( \mathbb{T} + \mathbb{V} - E \mathbb{1} \right)$) or a cylindrical wave $i(\alpha) H^{(+)}_{0}(k \rho)$ which is a solution of $\left( \mathbb{T} - E \mathbb{1} \right)$. The solution vector $v$ contains the expansion coefficients of the partial Faddeev component $f(\rho,\alpha)$. This Faddeev component consists further of two parts $f_a(\rho,\alpha)$ with $a=1,2$ for the two singlet and triplet channels.

\begin{figure}[!h]
    \centering
    \includegraphics[width=\textwidth]{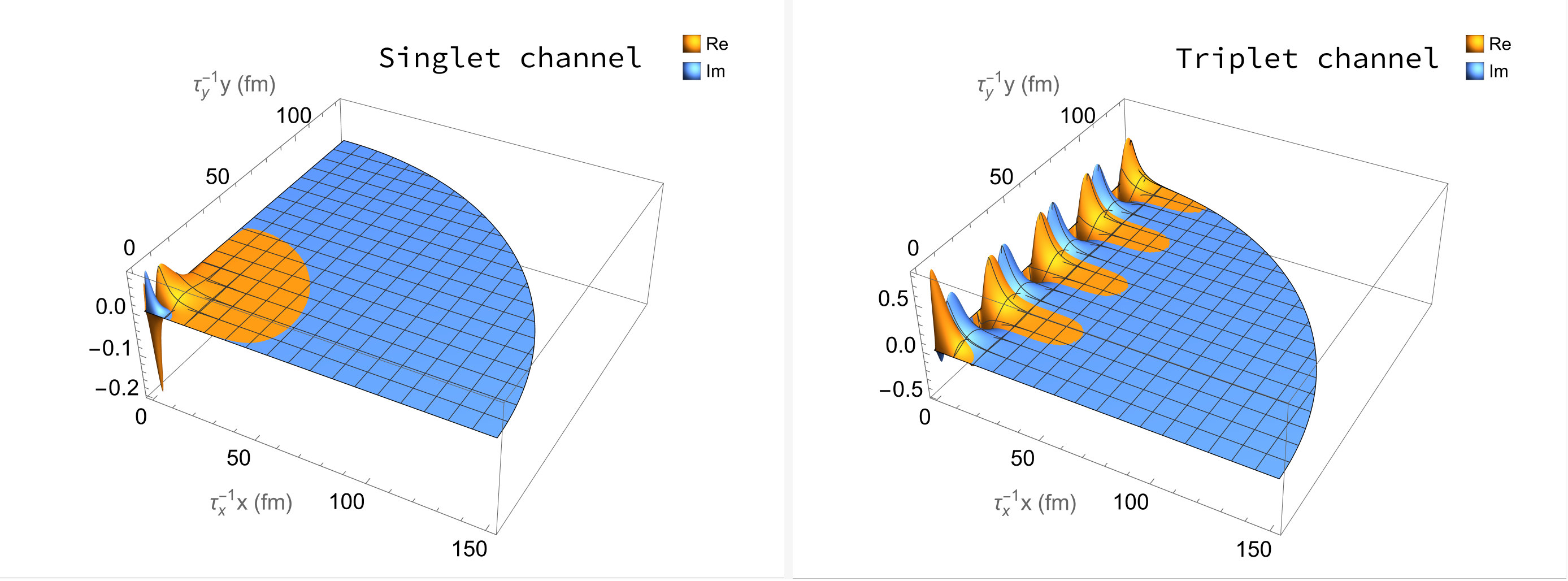}
    \caption{Calculated partial Faddeev components for $nd$ scattering, below breakup, at $E = -1$ MeV.}
    \label{fig:BelowComps}
\end{figure}

We show in figure~\eqref{fig:BelowComps} typical partial Faddeev components below breakup. Components are shown as a function of unscaled Jacobi distances $\tau_x^{-1}x$ and $\tau_y^{-1}y$ expressed in fm. At negative energies, the singlet channel is closed. The corresponding partial component is non-negligible only in the inner region where both channels are coupled. It vanishes exponentially elsewhere. The triplet component is seen to behave as a bound plane wave asymptotically, in the region $x$ finite and $y \to \infty$.

\begin{figure}[!h]
    \centering
    \includegraphics[width=\textwidth]{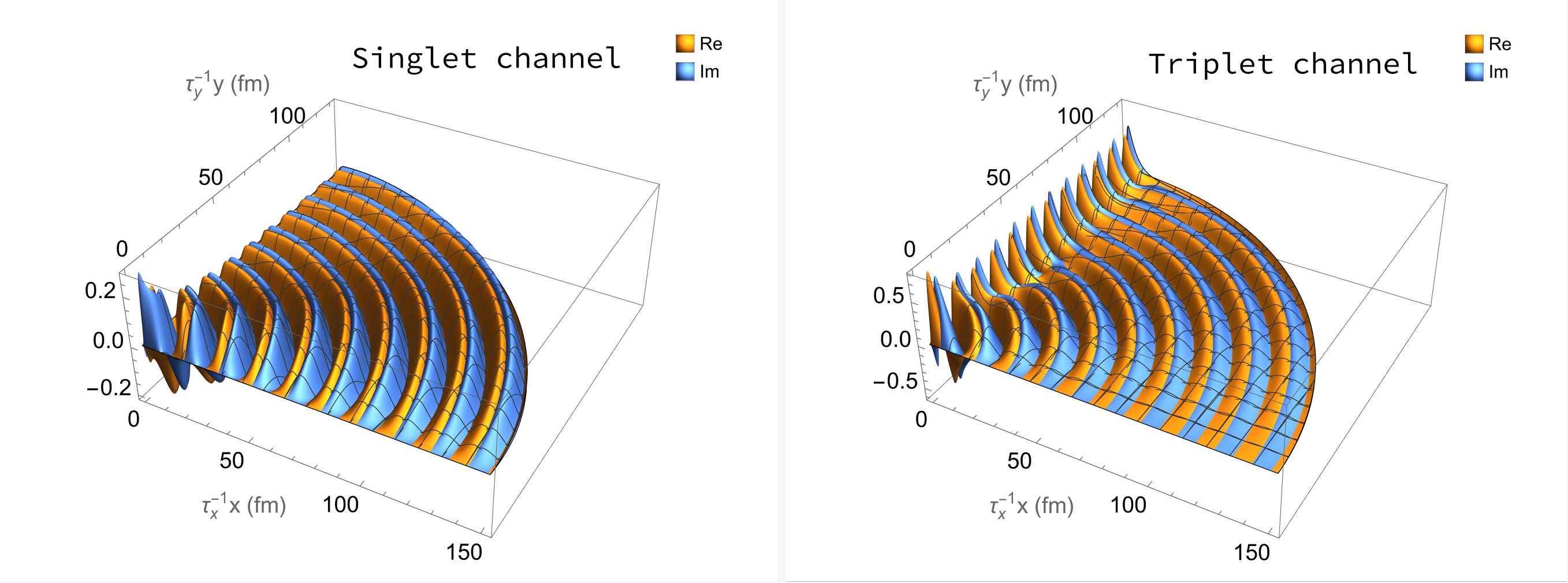}
    \caption{Calculated partial Faddeev components for $nd$ scattering, above breakup, at $E = 7.17$ MeV. This energy corresponds to $E_{lab} = 14.1$ MeV from~\cite{friar1995benchmark}.}
    \label{fig:AboveComps}
\end{figure}

We show in figure~\eqref{fig:AboveComps} typical partial Faddeev components above breakup. We have chosen $E \approx 7.17$ MeV which corresponds to an energy in the laboratory frame $E_{lab} = 14.1$ MeV which is one of the energy for which benchmark data are quoted in~\cite{friar1995benchmark}. The two energies are related by 
\begin{equation}
    \label{eq:elab}
    E = E_d + \frac23 E_{lab}   
\end{equation}
We can clearly see that the singlet component behaves as a cylindrical wave asymptotically. The triplet component have both a bound plane wave and a cylindrical wave behaviour.

\begin{figure}[!h]
    \centering
    \includegraphics[width=0.4\textwidth]{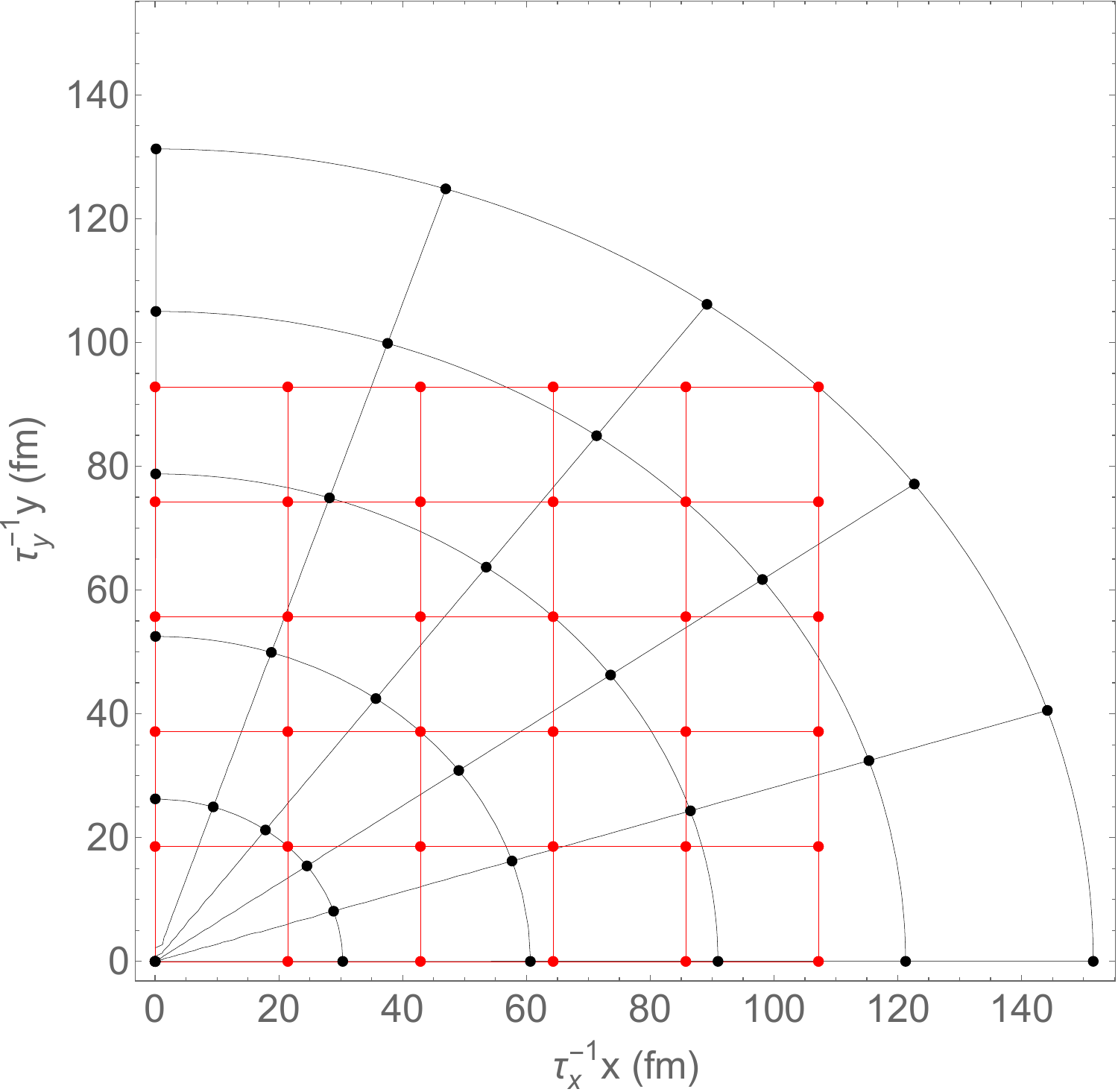}
    \caption{Schematics of the resampling process. The Faddeev components are calculated on a polar grid (black grid) as $f(\rho,\alpha)$. They are resampled on a cartesian grid (red grid) as $f(x,y)$ in order to extract the bound plane wave part from them.}
    \label{fig:resampling}
\end{figure}

In order to extract the bound plane wave part from the partial component $f_a(\rho,\alpha)$, this function is resampled as $f_a(x,y)$ on a cartesian grid inscribed in the polar grid. We schematically illustrate this process in figure~\eqref{fig:resampling}.

In its most general form, the function $f_a(x,y)$ behaves asymptotically as a linear combination of regular $\varphi_d(x) j(q y)$ and outgoing $\varphi_d(x) h^{(+)}(q y)$ bound plane waves as well as regular $J_{0}(k \rho)$ and outgoing $H_{0}^{(+)}(k\rho)$ cylindrical waves. Its projection on the deuteron bound state $\langle \varphi_d(x) | f_a(x,y) \rangle \equiv g(y)$ can therefore be written as 
\begin{align}
    g(y) &\approx c_{1} j(q y) + c_{2} h^{+}(q y) + c_{3} \mathfrak{j}(y) + c_{4} \mathfrak{h}^{(+)}(y) \\
    \mathfrak{j}(y) &= \left\langle \varphi_d(x) \Big\lvert i\left(\text{arctan}(y/x)\right)J_{0}\left(k \sqrt{x^2+y^2}\right) \right\rangle \\
    \mathfrak{h}^{(+)}(y) &= \left\langle \varphi_d(x) \Big\lvert i\left(\text{arctan}(y/x)\right)H_{0}^{(+)}\left(k \sqrt{x^2+y^2}\right) \right\rangle
\end{align}
where the coefficients $c_{1}$, $c_{2}$, $c_{3}$ and $c_{4}$ are unknowns which we determine through a non linear fit of the function $g(y)$ over a range of $y$ values. The two coefficients $c_{1}$ and $c_{2}$ will eventually define the elastic scattering amplitude. Since we have imposed a boundary condition pertaining to an outgoing cylindrical wave on the scattered Faddeev component, the coefficient $c_{3}$ should be zero which we indeed verify through the fit. Note that in the definition of the function $\mathfrak{h}^{(+)}(y)$ the unknown breakup amplitude has been replaced by the incoming function $i(\alpha)$. This approximation is justified a posteriori through the excellent quality of the fit. As a consequence, the coefficient $c_{4}$ has no physical meaning.
Having extracted the coefficients $c_{1}$ and $c_{2}$, we can subtract this bound plane wave part from the scattered Faddeev component. In its most general form, this difference behaves asymptotically as 
\begin{equation}
    f_{a}(\rho,\alpha) - c_{1} \varphi_d(\rho \cos \alpha) j(q \rho \sin \alpha) - c_{2} \varphi_d(\rho \cos \alpha) h^{(+)}(q \rho \sin \alpha) \approx C(\alpha)J_{0}(k\rho) + T_{ij}(\alpha)H_{0}^{(+)}(k\rho)
\end{equation}
where the functions $C(\alpha)$ and $T_{ij}(\alpha)$ are unknowns which we again determine through a non linear fit at each angle $\alpha$. As before, through the application of the boundary condition on the scattered Faddeev component the function $C(\alpha)$ should be zero which we indeed verify through the fit. Finally, the function $T_{ij}(\alpha)$ is the sought-after breakup amplitude, the relevant $\mathbf{T}$ matrix element appearing in eq.~\eqref{scattEq}.

Throughout these fitting procedures, it is important to include in the fit the regular parts and to verify that the corresponding coefficients ($c_{3}$ and $C(\alpha)$ above) are indeed zero. This serves as a numerical test on the quality of the calculated Faddev component.



The total dimension of the operators in equations~\eqref{operatorsEq} is $N=n_c n_a n_{\alpha} n_{\rho}$ where $n_c=1$ is the number of Faddeev components, $n_a = 2$ the number of channels, $n_{\alpha}$ the number of collocation points in $\alpha$ and $n_{\rho}$ the number of collocation points in $\rho$. In our application, we have $N = 256{\scriptstyle ,}000$ for a density around $0.1$\% ; the resolution of the linear system therefore cannot be done by a direct method. We have use the iterative GMRES~\cite{saad1986gmres} method in conjunction with a preconditioner consisting of $\left( \mathbb{T} - E \mathbb{1} \right)^{-1}$. Indeed, the operator $\left( \mathbb{T} - E \mathbb{1} \right)$ can be expressed as a sum of two Kronecker products \textit{i.e.} $\left( \mathbb{T} - E \mathbb{1} \right) = \mathbb{A}_1 \otimes \mathbb{B}_1 + \mathbb{A}_2 \otimes \mathbb{B}_2$ where the operators $\mathbb{A}$ are of dimension $n_{\rho}=500$ and the operators $\mathbb{B}$ of dimension $n_c n_a n_{\alpha}=512$. This allows to express the inverse of $\left( \mathbb{T} - E \mathbb{1} \right)$ through the evaluation of inverses and diagonalizations of those smaller operators (see for instance~\cite{schellingerhout1989configuration,lazauskas2019faddeev}). The iterative method with this preconditioner then converges in a few iterations.

\subsection{Elastic n-d scattering}

\begin{figure}[!h]
    \centering
    \includegraphics[width=\textwidth]{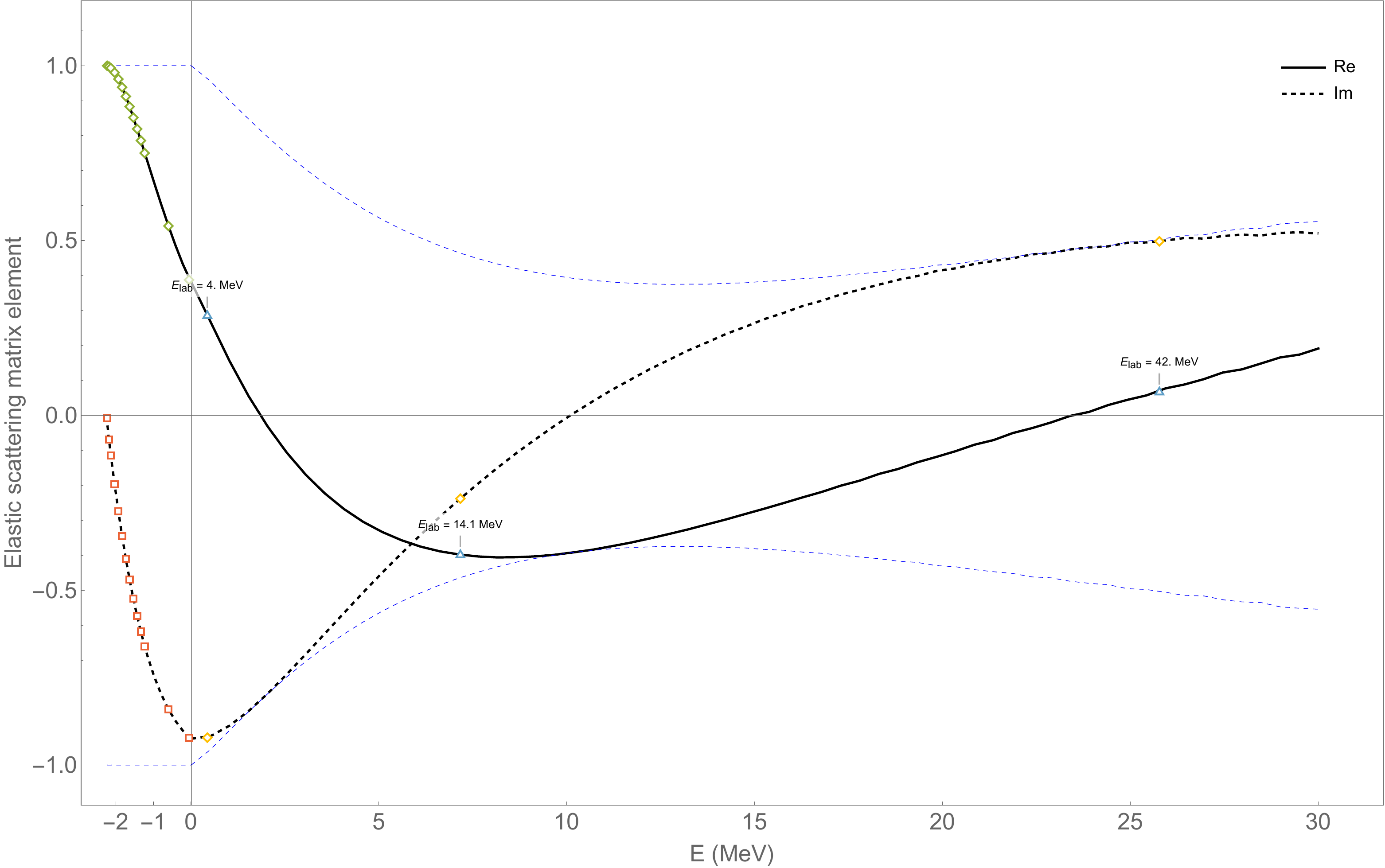}
    \caption{Elastic scattering matric element $S_{11}$ (black plain and dashed curves) for the doublet $J^{\Pi} = 1/2^{+}$ as a function of the center of mass energy $E$. Also shown is the envelope $\pm |S_{11}|$ (blue dashed curves). Data below breakup are from~\cite{chen1989low}. Data above breakup are from~\cite{friar1990benchmark,friar1995benchmark} and are tagged according to their energy in the laboratory frame (the relation between the two energies is given by eq.~\eqref{eq:elab}).} 
    \label{fig:elastic}
\end{figure}

We start by showing our results for the elastic $n$-$d$ scattering, which correspond to the element $S_{11}$ from equation~\eqref{scatteringMatrix}, in figure~\eqref{fig:elastic}. Calculations are done starting from the deuteron bound state energy $E_d$ up to around $E \approx 30$ MeV. This range is chosen to cover the energies for which benchmark data exists above breakup ; those energies are quoted in terms of the energy in the laboratory frame $E_{lab}$ (see equation~\eqref{eq:elab}).

Below breakup, the benchmark data are taken from reference~\cite{chen1989low} which uses a theoretical methods similar to ours \emph{i.e.} configuration-space Faddeev calculations using cubic splines orthogonal collocation.

Above breakup, the benchmark data are taken from references~\cite{friar1990benchmark,friar1995benchmark} and correspond to a collection of different theoretical methods both in configuration-space and momentum-space. The agreement between our calculations and the various benchmark data, both below and above breakup, is very good. This agreement validates our method of resampling the Faddeev components on a cartesian grid in order to extract the $S_{11}$ matrix element from them.

\subsection{Breakup amplitudes}

\begin{figure}[!h]
    \centering
    \includegraphics[width=\textwidth]{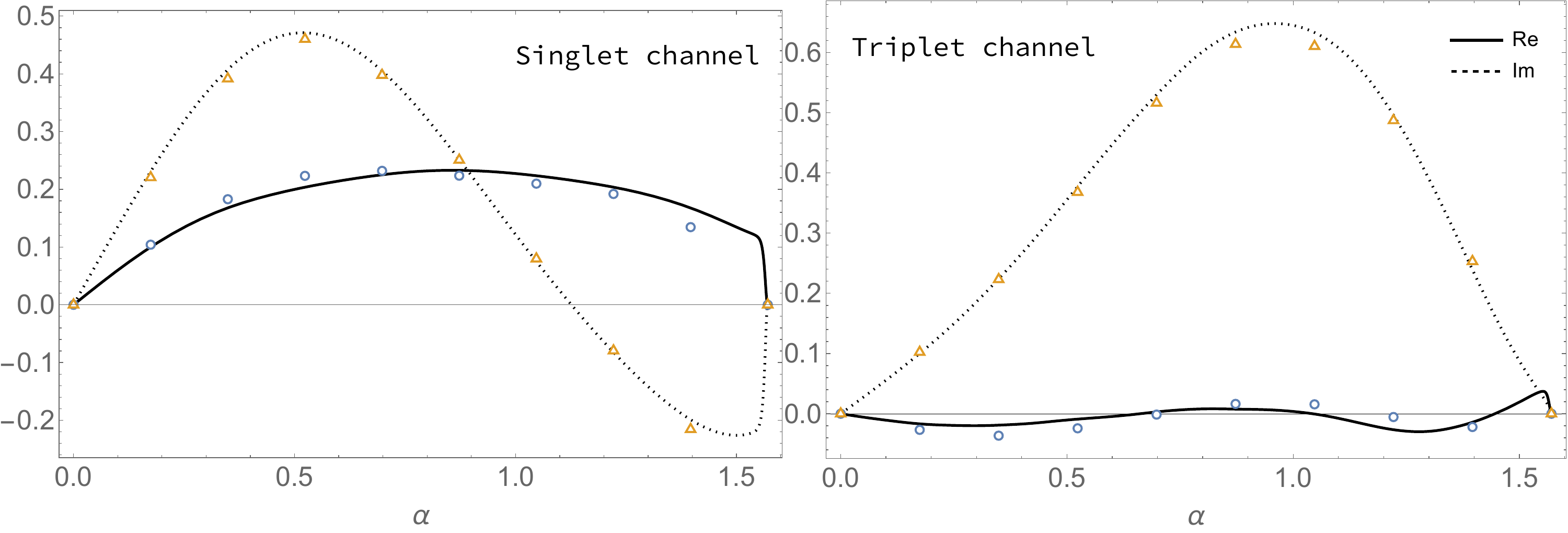}
    \caption{Breakup amplitudes $\sqrt{\frac{2}{\pi}} \frac{e^{i \pi / 4}}{2i} S_{12}(\alpha)$ and $\sqrt{\frac{2}{\pi}} \frac{e^{i \pi / 4}}{2i} S_{13}(\alpha)$ in the singlet and triplet channels at $E_{lab} = 14.1 $ MeV for the doublet system $J^{\Pi} = 1/2^{+}$. Benchmark data are from reference~\cite{friar1995benchmark}.}
    \label{fig:breakAmps14}
\end{figure}

We then show our results for the breakup amplitudes, which correspond to the matrix elements $S_{12}(\alpha)$ and $S_{13}(\alpha)$ from equation~\eqref{scatteringMatrix}, in figure~\eqref{fig:breakAmps14}.

The benchmark data are taken from reference~\cite{friar1995benchmark}. In order to facilitate the comparison between our calculations and the benchmark data, our breakup amplitudes are multiplied by $\sqrt{\frac{2}{\pi}} \frac{e^{i \pi / 4}}{2i}$. Indeed, in reference~\cite{friar1995benchmark} a matrix element of $\mathbf{T}$ instead of $\mathbf{S}$ is presented and this matrix element is expressed with respect to the asymptotic function $\frac{e^{i k \rho}}{\sqrt{k \rho}}$ instead of $H_{0}^{(+)}(k \rho)$. The agreement between our breakup amplitudes and the benchmark data is very satisfactory. We have also compared our breakup amplitudes at $E_{lab} = 4$ MeV and $E_{lab} = 42$ MeV with the same level of agreement.
We can see in fig.~\eqref{fig:breakAmps14} what we have discussed in section~\ref{sec:reaching} : the sharp drop near $\alpha \approx \pi / 2$ results from the fact that this region is not converged given our grid in $\rho$. One would have to do a calculation for a larger value of $\rho$ in order to have breakup amplitudes converged for all values of $\alpha$.

\subsection{Three-body recombination}

Nuclear reactions involving three particles are exceptionally rare so there is no data on the three-body recombination reaction $n+n+p \to n+d$ to the best of our knowledge. Nevertheless, our formalism leads to the full scattering matrix so it calculates the probabilities for this process. According to equation~\eqref{scatteringMatrix} the three-body recombination processes corresponds to the matrix elements $S_{21}$ and $S_{31}$. They differ by the two incoming functions $i_1(\alpha)$ and $i_2(\alpha)$ which characterize the incoming $n+n+p$ state (see equation~\eqref{scattEq}). Following the discussion in section~\ref{subs:doubleCont}, we take for these incoming functions the first two Delves functions $D_{\nu}^{(\ell,\lambda)}(\alpha)$. In our case where $\ell = \lambda = 0$ we have $D_{\nu}^{(0,0)}(\alpha) \propto \sin \left(2(n+1) \alpha\right)$ with $n=0,1,2\ldots$

\begin{figure}[!h]
    \centering
    \includegraphics[width=\textwidth]{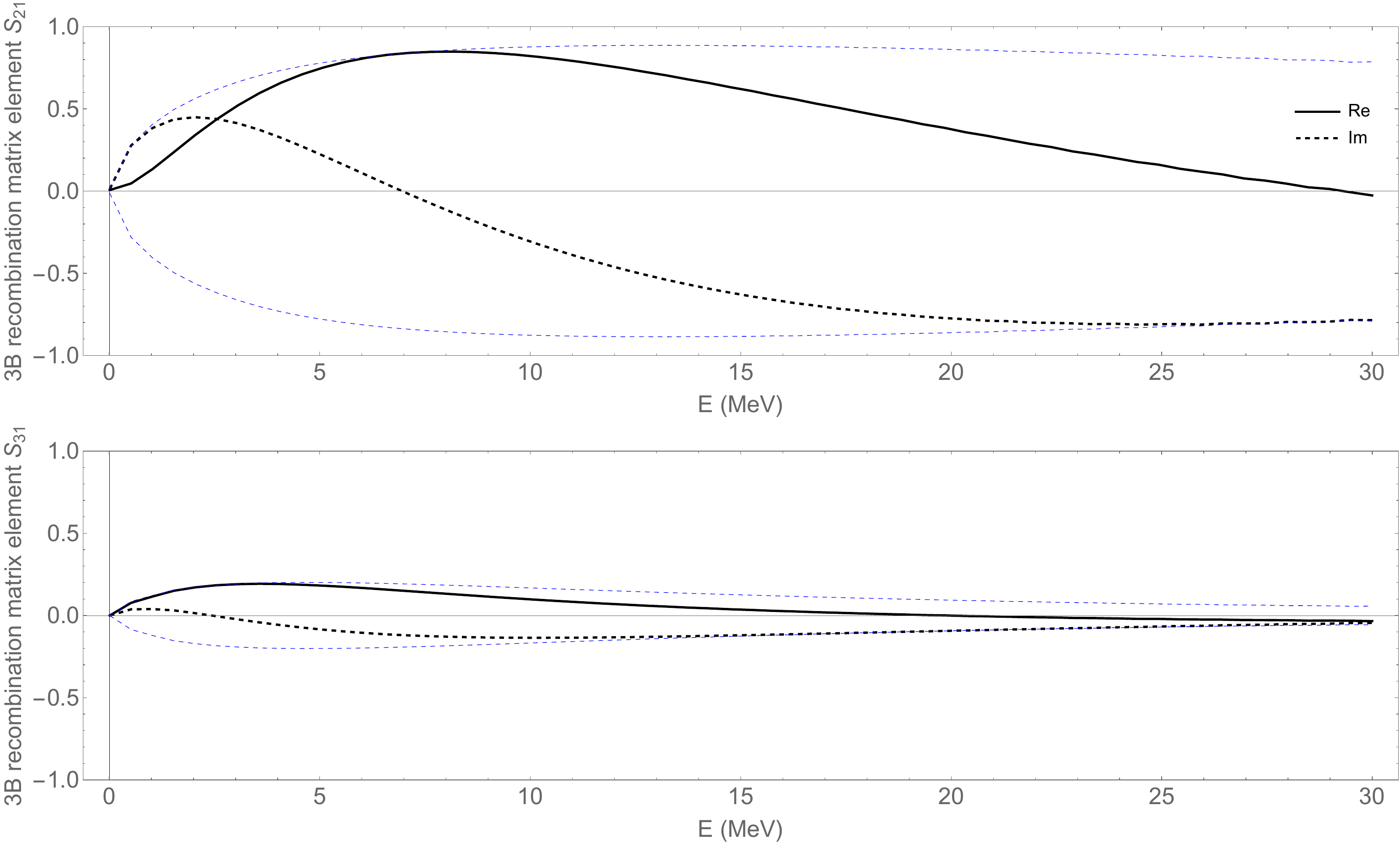}
    \caption{Three-body recombination matrix elements $S_{21}$ and $S_{31}$ from initial states consisting of the first two Delves functions $\sin 2\alpha$ and $\sin 4\alpha$ respectively. The envelope is also shown in dashed blue lines.}
    \label{fig:tbr}
\end{figure}

The calculated three-body recombination matrix elements $S_{21}$ and $S_{31}$ as a function of the energy are shown figure~\eqref{fig:tbr}.

\subsection{Elastic n-n-p scattering} 
\label{sub:elastic_nnp}

Finally, we present some results for the elastic $n+n+p \to n+n+p$ scattering. The initial $n+n+p$ is taken to be describe by the function $i_{1}(\alpha) \propto \sin 2 \alpha$ (see the discussion above for three body recombination processes). This process corresponds to the matrix elements $S_{22}(\alpha)$ and $S_{23}(\alpha)$ in equation~\eqref{scatteringMatrix}.
\begin{figure}[!h]
    \centering
    \includegraphics[width=\textwidth]{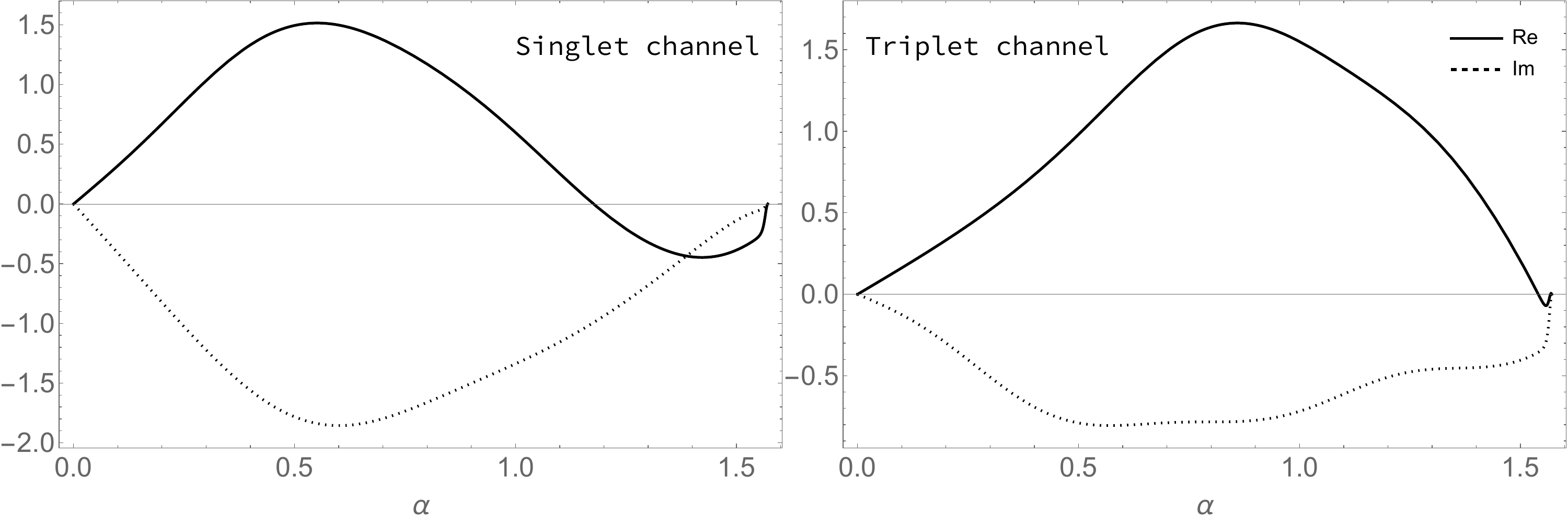}
    \caption{Amplitudes for elastic $n+n+p \to n+n+p$ scattering at $E=0.43$ MeV (which corresponds to $E_{lab} = 4$ MeV).}
    \label{fig:nnpElas}
\end{figure}
The calculated elastic $n+n+p$ scattering matrix elements $S_{22}(\alpha)$ and $S_{23}(\alpha)$ for an energy $E=0.43$ MeV are shown figure~\eqref{fig:nnpElas}.


\subsection{Properties of the scattering matrix}

The scattering matrix in equation~\eqref{scatteringMatrix} is very compact and describe all processes to and from a single continuum channel or a double continuum one. However, its matrix elements are both regular complex numbers (for processes ending in a single continuum state) and complex functions of the polar angle $\alpha$ (for processes ending in a double continuum state). In this section, we present how to retrieve usual properties like unitarity and reciprocity from such a scattering matrix. First, we define a matrix of incident functions $\boldsymbol{\mathfrak{I}}$ as
\begin{equation}
    \boldsymbol{\mathfrak{I}} = \left( \begin{array}{c|cc}
        1 & 0  & 0 \\
         \hline
        0 & i_{1}(\alpha) & 0 \\
        0 & 0 & i_{2}(\alpha)
    \end{array} \right)
\end{equation}
It is simply the matrix in front of the regular functions in equation~\eqref{scattEq} and collects the functions describing the incident double continuum states.

Then, we define a custom binary operation $\langle \cdot,\cdot \rangle$ which will act on the scattering matrix elements. Let $\gamma$, $\delta$ be complex numbers and let $f(\alpha)$, $g(\alpha)$ be complex functions of the polar angle $\alpha$. The binary operation $\langle \cdot,\cdot \rangle$ is defined as
\begin{subequations}
    \begin{align}
        \langle \gamma,\delta \rangle &= \gamma \delta \\
        \langle \gamma,g      \rangle &= \gamma \int d \alpha \, \left(g(\alpha) + \mathcal{J}\left[ g \right](\alpha)\right) \\
        \langle f,\delta      \rangle &= \delta \int d \alpha f(\alpha) \\
        \langle f,g           \rangle &= \int d \alpha f(\alpha)\left(g(\alpha) + \mathcal{J}\left[ g \right](\alpha)\right)
    \end{align}
\end{subequations}
In addition to reducing to the usual multiplication, it performs an integration over all polar angle and importantly implements the Jacobi transform with respect to the second argument (for the sake of clarity, we dropped the subscripts and superscripts of the Jacobi transform ; it has to be understood as a sum over all different components and all partial waves).

Finally, we define a $*$ operation between two matrices. This operation corresponds to the usual matrix product (\emph{i.e.} a dot product between lines of the first matrix and columns of the second one) where the multiplication is replaced by our custom binary operation $\langle \cdot,\cdot \rangle$. Thanks to this $*$ operation, we are able to obtain familiar and compact formula both for the unitarity and the reciprocity of the scattering matrix $\mathbf{S}$.

\subsubsection{Unitarity}

The unitarity condition for $\mathbf{S}$ reads
\begin{equation}
    \mathbf{S} * \mathbf{S}^{\dag} = \boldsymbol{\mathfrak{I}} * \boldsymbol{\mathfrak{I}}^{\dag}
\end{equation}
The above expression equates outgoing probability flux (the left hand side) to incoming flux (the right hand side). Further, we can always choose the functions $i(\alpha)$ in $\boldsymbol{\mathfrak{I}}$ to have unit incoming flux with respect to $\langle \cdot,\cdot \rangle$ to obtain
\begin{equation}
    \mathbf{S} * \mathbf{S}^{\dag} = \mathbf{1}
\end{equation}
We note however that a unit flux with respect to $\langle \cdot,\cdot \rangle$ does not mean that the functions $i(\alpha)$ are normalized in the usual way \emph{i.e.}
\begin{equation}
    \langle i , i \rangle = 1 \centernot\implies \langle i | i \rangle = 1
\end{equation}
because of the contributions from the Jacobi transform.

We can then define a defect from unitarity $\eta_{U}$ as
\begin{equation}
    \eta_{U} = \left\| \mathbf{1} - \mathbf{S} * \mathbf{S}^{\dag} \right\|
\end{equation}
where $\| \cdot \|$ is a matrix norm. We stress that $\eta_{U}$ measures not only the conservation of probabilities of the various scattering processes (as the diagonal elements of $\mathbf{S} * \mathbf{S}^{\dag}$ being equal to one) but also the degree of orthogonality of the asymptotic channels (as the off-diagonal elements of $\mathbf{S} * \mathbf{S}^{\dag}$ being equal to zero).

\subsubsection{Reciprocity}

In order to verify reciprocity, the breakup amplitudes $S_{12}(\alpha)$ and $S_{13}(\alpha)$ have to be projected with respect to $\langle \cdot,\cdot \rangle$ onto the incoming functions $i_{1}(\alpha)$ and $i_{2}(\alpha)$ in $\boldsymbol{\mathfrak{I}}$. This can be perform in a compact way using the $*$ operation and thus we define
\begin{equation}
    \boldsymbol{\mathcal{S}} = \boldsymbol{\mathfrak{I}} * \mathbf{S}^{\text{T}}
\end{equation}
and the reciprocity property of $\mathbf{S}$ means that $\boldsymbol{\mathcal{S}}$ must be symmetric. We then define a defect from reciprocity as
\begin{equation}
    \eta_{R} = \frac{\left\| \boldsymbol{\mathcal{S}} - \boldsymbol{\mathcal{S}}^{\text{T}} \right\|}{\left\| \boldsymbol{\mathcal{S}} + \boldsymbol{\mathcal{S}}^{\text{T}} \right\|}
\end{equation}

\begin{figure}[!h]
    \centering
    \includegraphics[width=\textwidth]{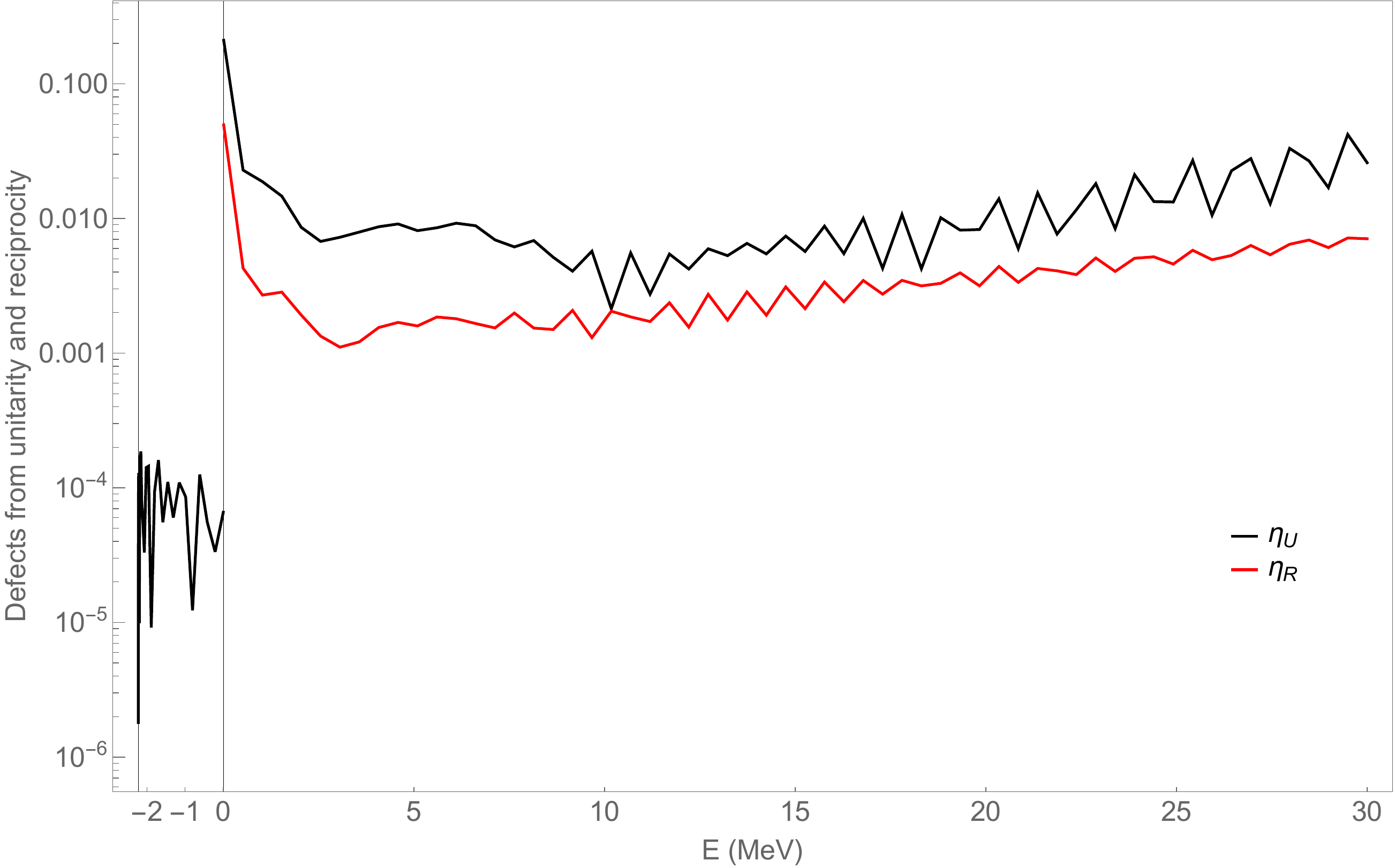}
    \caption{Defects from unitarity and reciprocity of the calculated scattering matrix as a function of the energy.}
    \label{fig:defects}
\end{figure}

We show in figure~\eqref{fig:defects} the defects from unitarity and reciprocity of the calculated scattering matrix as a function of the energy. Several features are worth mentioning.
At negative energy where the $\mathbf{S}$ matrix reduces to the single element $S_{11}$ the quality of the calculations is very satisfactory. The square modulus of $S_{11}$ only differing from one at a level of $10^{-5} - 10^{-4}$.

At positive energy, there is an overall degradation in the accuracy of the calculation. Of notice is the poor values of both $\eta_{U}$ and $\eta_{R}$ at small positive energies. This is probably due to the fact that our calculation is done for a fixed grid designed to represent distances of the order of $100$ fm. Since the asymptotic region in the double continuum channels involves interferences from the cylindrical waves from different Faddeev components, one needs to use a much larger grid in $\rho$ when the wavenumber $k = \sqrt{E}$ in $H^{(+)}_{0}(k \rho)$ is small in order to describe several wavelengths of the cylindrical waves.

Another feature are the apparent oscillations in the defects as a function of the energy (note however that because of the logarithmic scale, those are small in amplitude). This is probably due to the fact that, as we mentioned in section~\ref{subs:numerical}, in imposing a boundary condition adapted to an outgoing cylindrical wave this boundary condition is not adapted to the bound plane wave part of the Faddeev component. As a function of the energy, the applied boundary condition will be periodically rather well adapted to an outgoing bound plane wave though.

In the end, the defects are on a level of $10^{-3} - 10^{-2}$ on the full range of energies which is in fact comparable to our previous calculation~\cite{guerout2024calculation} and stays satisfactory.

\section{Conclusions}

In this work, we have presented a simple method to perform scattering Faddeev calculations in the double continuum based on a resampling of the calculated polar wavefunction on a cartesian grid. We have apply successfully our method on the benchmark neutron-deuteron $J=1/2$ scattering. Note that we have also performed scattering calculations for the quartet $J=3/2$ with the same level of agreement to benchmark data.

We have unify both the single and double continuum channels in a unique scattering matrix and shown how to retrieve its usual properties.

\section*{Declaration of interests}

The authors do not work for, advise, own shares in, or receive funds from any organisation that could benefit from this article, and have declared no affiliation other than their research organisations.

\bibliographystyle{unsrt}
\bibliography{refs}

@article{delves1960tertiary,
  title={Tertiary and general-order collisions (II)},
  author={Delves, LM},
  journal={Nuclear Physics},
  volume={20},
  pages={275--308},
  year={1960},
  publisher={Elsevier}
}

@article{guerout2024calculation,
  title={Calculation of bound and continuum states of the Ne3 van der Waals trimer},
  author={Gu{\'e}rout, Romain},
  journal={Journal of Physics B: Atomic, Molecular and Optical Physics},
  volume={58},
  number={1},
  pages={015201},
  year={2024},
  publisher={IOP Publishing}
}

@article{glockle1992boundary,
  title={Boundary conditions for three-body scattering in configuration space},
  author={Gl{\"o}ckle, W and Payne, GL},
  journal={Physical Review C},
  volume={45},
  number={3},
  pages={974},
  year={1992},
  publisher={APS}
}

@article{belov2013asymptotic,
  title={Asymptotic method for determining the amplitude for three-particle breakup: Neutron-deuteron scattering},
  author={Belov, PA and Yakovlev, SL},
  journal={Physics of Atomic Nuclei},
  volume={76},
  number={2},
  pages={126--138},
  year={2013},
  publisher={Springer}
}

@article{friar1995benchmark,
  title={Benchmark solutions for n-d breakup amplitudes},
  author={Friar, James Lewis and Payne, GL and Gl{\"o}ckle, W and H{\"u}ber, D and Wita{\l}a, H},
  journal={Physical Review C},
  volume={51},
  number={5},
  pages={2356},
  year={1995},
  publisher={APS}
}

@article{chen1989low,
  title={Low-energy nucleon-deuteron scattering},
  author={Chen, CR and Payne, GL and Friar, James Lewis and Gibson, Benjamin F},
  journal={Physical Review C},
  volume={39},
  number={4},
  pages={1261},
  year={1989},
  publisher={APS}
}

@article{roudnev2011automatic,
  title={Automatic grid construction for few-body quantum-mechanical calculations},
  author={Roudnev, Vladimir and Cavagnero, Michael},
  journal={Computer Physics Communications},
  volume={182},
  number={10},
  pages={2099--2106},
  year={2011},
  publisher={Elsevier}
}

@article{lazauskas2019faddeev,
  title={The faddeev--yakubovsky symphony},
  author={Lazauskas, Rimantas and Carbonell, Jaume},
  journal={Few-Body Systems},
  volume={60},
  number={4},
  pages={62},
  year={2019},
  publisher={Springer}
}

@article{saad1986gmres,
  title={GMRES: A generalized minimal residual algorithm for solving nonsymmetric linear systems},
  author={Saad, Youcef and Schultz, Martin H},
  journal={SIAM Journal on scientific and statistical computing},
  volume={7},
  number={3},
  pages={856--869},
  year={1986},
  publisher={SIAM}
}

@book{edmonds1996angular,
  title={Angular momentum in quantum mechanics},
  author={Edmonds, Alan Robert},
  volume={4},
  year={1996},
  publisher={Princeton university press}
}

@article{nadal2011method,
  title={A method to compute probability current in generic coordinates},
  author={Nadal-Ferret, Marc and Gelabert, Ricard and Moreno, Miquel and Lluch, Jos{\'e} M},
  journal={The Journal of chemical physics},
  volume={134},
  number={7},
  year={2011},
  publisher={AIP Publishing}
}

@article{friar1990benchmark,
  title={Benchmark solutions for a model three-nucleon scattering problem},
  author={Friar, James Lewis and Gibson, Benjamin F and Berthold, G and Gl{\"o}ckle, W and Cornelius, Th and Witala, H and Haidenbauer, J and Koike, Y and Payne, GL and Tjon, JA and others},
  journal={Physical Review C},
  volume={42},
  number={5},
  pages={1838},
  year={1990},
  publisher={APS}
}

@article{schellingerhout1989configuration,
  title={Configuration-space Faddeev calculations: Supercomputer accuracy on a personal computer},
  author={Schellingerhout, NW and Kok, LP and Bosveld, GD},
  journal={Physical Review A},
  volume={40},
  number={10},
  pages={5568},
  year={1989},
  publisher={APS}
}

@article{kolganova1998three,
  title={Three-body configuration space calculations with hard-core potentials},
  author={Kolganova, EA and Motovilov, AK and Sofianos, SA},
  journal={Journal of Physics B: Atomic, Molecular and Optical Physics},
  volume={31},
  number={6},
  pages={1279},
  year={1998},
  publisher={IOP Publishing}
}

@article{merkuriev1976three,
  title={Three-body scattering in configuration space},
  author={Merkuriev, SP and Gignoux, C and Laverne, A},
  journal={Annals of physics},
  volume={99},
  number={1},
  pages={30--71},
  year={1976},
  publisher={Elsevier}
}

@article{fadeev1961scattering,
  title={Scattering theory for a three-particle system},
  author={Fadeev, LD},
  journal={Sov. Phys.-JETP},
  volume={12},
  pages={1014--1019},
  year={1961}
}

\end{document}